\documentclass[a4paper,fleqn]{cas-dc}
\usepackage[authoryear]{natbib}

\usepackage{makecell}
\usepackage{amsmath}
\usepackage{gensymb}
\usepackage{multirow}
\usepackage[normalem]{ulem}
\useunder{\uline}{\ul}{}
\usepackage{geometry}
\usepackage{pdflscape}
\usepackage{longtable,booktabs}
\usepackage{supertabular}
\usepackage{tabu}
\usepackage{afterpage,caption}
\usepackage{xcolor}
\usepackage{ragged2e}
\usepackage{url}

\usepackage[switch]{lineno}
\usepackage{draftwatermark}
\SetWatermarkLightness{0.9}
\SetWatermarkScale{1}

\newcommand{\hide}[1]{}

\usepackage{multicol}
\usepackage{glossaries}
\usepackage{balance}

\begin{document}


\let\WriteBookmarks\relax
\def\floatpagepagefraction{1}
\def\textpagefraction{.001}

\shorttitle{Wastewater-based epidemiology survey}

\shortauthors{Chen et~al.}
\title[mode=title]{Wastewater-based Epidemiology for COVID-19 Surveillance and Beyond: A Survey}

\author[1]{Chen Chen}[orcid=0000-0002-7423-0090]
\cormark[1]
\ead{chenannie45@gmail.com}
\affiliation[1]{organization={Department of Computer Science, University of Virginia},
city={Charlottesville},
postcode={22904},
country = {United States}}

\author[1]{Yunfan Wang}
\ead{yunf.wang@outlook.com}

\author[2]{Gursharn Kaur}
\ead{fug3aj@virginia.edu}
\affiliation[2]{organization={Biocomplexity Institute and Initiative, University of Virginia},
city={Charlottesville},
postcode={22904},
country={United States}}

\author[2]{Aniruddha Adiga}
\ead{aa5dw@virginia.edu}

\author[2]{Baltazar Espinoza}
\ead{be8dq@virginia.edu}

\author[2]{Srinivasan Venkatramanan}
\ead{sv8nv@virginia.edu}

\author[2]{Andrew Warren}
\ead{sasw3xp@virginia.edu}

\author[2]{Bryan Lewis}
\ead{bl4zc@virginia.edu}

\author[3]{Justin Crow}
\ead{justin.crow@vdh.virginia.gov}
\affiliation[3]{organization={Virginia Department of Health},
city={Richmond},
postcode={23219},
country={United States}}

\author[3]{Rekha Singh}
\ead{rekha.singh@vdh.virginia.gov}

\author[4]{Alexandra Lorentz}
\ead{alexandra.lorentz@dgs.virginia.gov}
\affiliation[4]{organization={Division of Consolidated Laboratory Services, Department of General Services},
city={Richmond},
postcode={23219},
country = {United States}}

\author[4]{Denise Toney}
\ead{denise.toney@dgs.virginia.gov}

\author[1,2]{Madhav Marathe}
\ead{marathe@virginia.edu}

\cortext[cor1]{Corresponding author}

\begin{abstract}
The pandemic of COVID-19 has imposed tremendous pressure on public health systems and social economic ecosystems over the past years. To alleviate its social impact, it is important to proactively track the prevalence of COVID-19 within communities. The traditional way to estimate the disease prevalence is to estimate from reported clinical test data or surveys. However, the coverage of clinical tests is often limited and the tests can be labor-intensive, requires reliable and timely results, and consistent diagnostic and reporting criteria. Recent studies revealed that patients who are diagnosed with COVID-19 often undergo fecal shedding of SARS-CoV-2 virus into wastewater, which makes wastewater-based epidemiology for COVID-19 surveillance a promising approach to complement traditional clinical testing. In this paper, we survey the existing literature regarding wastewater-based epidemiology for COVID-19 surveillance and summarize the current advances in the area. Specifically, we have covered the key aspects of wastewater sampling, sample testing, and presented a comprehensive and organized summary of wastewater data analytical methods. Finally, we provide the open challenges on current wastewater-based COVID-19 surveillance studies, aiming to encourage new ideas to advance the development of effective wastewater-based surveillance systems for general infectious diseases.

\end{abstract}






\maketitle

\section{Introduction}

The pandemic of COVID-19 has posed significant challenges to public health systems and the global economy, thereby urging the need for effective surveillance methods to monitor the prevalence of the disease within communities. Conventional surveillance methods are heavily dependent on clinical test data, such as positive test cases and hospitalizations. The inherent limitation of clinical data-based surveillance methods lies in their limited coverage, labor intensity, and data staleness due to prolonged test procedures. In order to estimate the prevalence of the disease and detect potential outbreaks in a more timely fashion, wastewater-based epidemiology (WBE\footnote{In this paper, our primary focus is on the surveillance perspective of WBE when using the term "WBE". All acronyms that appear more than once in the paper are summarized in Appendix Table~\ref{tab:acronyms}.}) surveillance has been identified as complementary to clinical methods.   

\begin{figure*}
    \centering
    \includegraphics[width=0.9\textwidth]{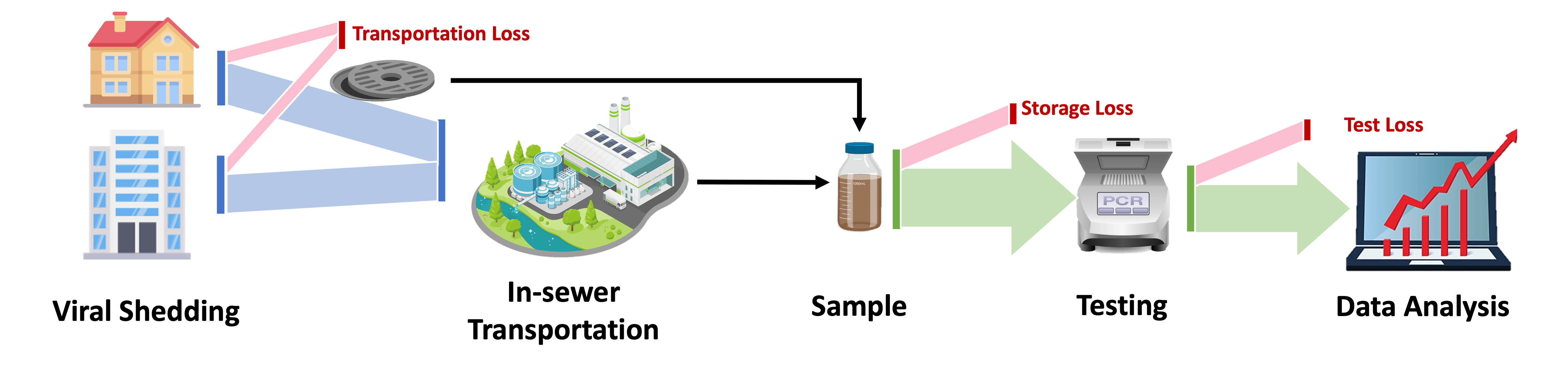}
    \caption{Overview of Wastewater-based Epidemiology Surveillance System.}
    \label{fig:ww_overview}
\end{figure*}

WBE has been successfully used for monitoring the use of pharmaceuticals ~\citep{bischel2015pathogens}, illicit drugs~\citep{zuccato2008estimating}, flu prevalence~\citep{heijnen2011surveillance}, and polio outbreaks~\citep{brouwer2018epidemiology}. Recent research suggests that monitoring the SARS-CoV-2 and other disease levels in wastewater can be a reliable way to understand the disease prevalence in addition to the clinical test results~\citep{safford2022wastewater}. Specifically, the wastewater samples can be collected from manholes in the targeted communities or from the wastewater treatment plants (WWTPs) in the sewersheds. The collected samples are then tested to quantify the concentration and the total load of the SARS-CoV-2 virus. The resulting viral concentration/load can be viewed as a comprehensive snapshot of disease prevalence within the community. By collectively analyzing the viral data from multiple timestamps, the trajectory of the disease may be estimated, which can be further used for trend projection. Figure~\ref{fig:ww_overview} shows the overview of the wastewater-based epidemic surveillance system.

While a promising tool, wastewater-based COVID-19 surveillance and beyond is subject to some key limitations and challenges. The first challenge is the variability in viral shedding rates. Specifically, individuals of different symptom severity and age groups may contribute virus to the sewage system at significantly different rates, thus making it hard to approximate the infected population from wastewater viral load. Second, the wastewater viral load may get underestimated due to dilution in the sewer system, in-sewer transportation loss, degradation of the virus, and also the test procedures used. Such loss is inevitable and could lead to missed cases or delayed alerts for outbreaks. On the other hand, the sewershed population, wastewater flow variations, and sample methods may also affect the representativeness of the viral level in the test sample to the disease prevalence of the entire community. Therefore, approximating the actual viral load that flows into the sewage system from degraded signals requires careful modeling and analysis. The last challenge is the integration of wastewater analysis with conventional surveillance results (e.g. reported cases, hospitalization). Wastewater-based surveillance data provides a comprehensive snapshot of disease prevalence within the whole community but with potentially considerable degradation. In contrast, conventional surveillance results are accurate but only cover a limited portion of the infected population. Effectively combining the two data sources can be problematic as the studied populations are not well aligned.

In this paper, we survey the current literature that encompasses critical facets of wastewater-based surveillance for COVID-19, including wastewater sampling techniques, sample testing methodologies, data analysis methods, available datasets at the global level, and the extension to other infectious diseases.
Furthermore, we highlight the ongoing challenges in the wastewater-based COVID-19 surveillance systems and hope to inspire continued innovation and development in the domain. It is worth mentioning that the data analytic methods for COVID-19 can be easily generalized to the surveillance tasks for other infectious diseases summarized in~\citet{kilaru2023wastewater}.

\hide{
\citet{sharara2021wastewater} describe the types of data that can be obtained through varying levels of WBE analysis, concrete plans for implementation, and public health actions that can be taken based on WBE surveillance data of infectious diseases, using recent and successful applications of WBE during the COVID-19 pandemic for illustration. describe how WBE is used as an early warning system,
an independent data stream to monitor disease spread and evaluate the effectiveness of public health interventions, and as a complement to genomic epidemiology.

\citet{polo2020making}  summarizes current knowledge and discusses the critical factors for implementing wastewater-based epidemiology of COVID-19. critical factors for WBE implementation (sampling, virus recovery and concentration, virus quantification, viral shedding rate and levels in ww, population normalization, ethical considerations)

\citet{shah2022wastewater} assess the performance of wastewater surveillance as early warning system of COVID-19 community transmission. Overall sample positivity was moderate at 29.2\% in all examined settings with the spike(S) gene having maximum rate of positive detections and nucleocapsid (N)gene being the most targeted.Wastewater signals preceded confirmed cases by up to 63days,with 13 studies reporting sample positivity before the first cases were detected in the community.At least 50 studies reported an association of viral load with community cases.While wastewater surveillance cannot replace large-scale diagnostic testing, it can complement clinical surveillance by providing early signs of potential transmission for more active public health responses. summarized sample collection and testing methodoloty in related studies. qualitative study on the association between viral load to cases. briefly mentioned the sequencing and modeling studies and risk assessment (uncertainty).

\citet{hamouda2021wastewater} conducted a comparison between the various studies with regards to sample concentration and virus quantification was conducted. Correlating the slope of the relationship between the number of gene copies vs. the cumulative number of infections normalized to the total population served with the average new cases,suggests that qPCR results could help estimating the number of new infections.The correlation is improved when a lag period was introduced to account for asymptomatic infections. factors affecting viral stability and RNA (decay factor) fate in wastewater, WBE framework; methods of detection and quantification of SARS-CoV-2 in wastewater.

\citet{ciannella2023recent} systematic review that organizes and discusses laboratory procedures for SARS-CoV-2 RNA quantification from a wastewater matrix,along with modeling techniques applied to the development of WBE for COVID-19 surveillance.The goal of this review is to present the current panorama of WBE operational aspects as well as to identify current challenges related to it. aspects of ww analysis for sars-cov-2 detection and quantification (pre-treatment, concentration, detection and quantification); correlation clinical test data to viral concentration in ww (ignores ww type, sample frequency, significance analysis, not comprehensive);  modeling of wbe for covid-19 surveillance (not clear taxonomy and comparison between different methods).

\citep{li2023correlation} correlations analysis (ignore the correlations type, not comprehensive).  identify the correlation between CRNA and various types of clinically confirmed case numbers, including prevalence and incidence rates. The impacts of environmental factors, WBE sampling design, and epidemiological conditions on the correlation were assessed for the same datasets. 
}

\noindent\textbf{Differences with Existing Surveys.} Existing surveys on waste\-water-based COVID-19 surveillance are predominantly focused on sampling methods, virus detection and quantification, and surveillance system design~\citep{sharara2021wastewater,polo2020making,shah2022wastewater,hamouda2021wastewater}. In~\citet{ciannella2023recent, li2023correlation}, the two surveys have covered the correlation analysis between viral concentration and clinical test results, but the studies are not comprehensive enough to cover all the critical aspects of the analysis (e.g., sample type, sample frequency, correlation metrics). To the best of our knowledge, this is a thorough survey that focuses on summarizing the state-of-the-art analytical methods used in  waste\-water-based COVID-19 surveillance and beyond. 

\noindent\textbf{Survey Structure.} The remainder of this survey is organized as follows, Section 2 and Section 3 briefly introduce the current advances in wastewater sampling and sample testing. Section 4 covers different aspects of wastewater analytic methods. Section 5 provides a comprehensive list of wastewater datasets for SARS-CoV-2 surveillance. Section 6 discusses the current limitations and challenges of wastewater-based COVID-19 surveillance systems, and Section 7 concludes the survey.

\section{Literature Collection and Organization}
Following the guidance of the Preferred Reporting Items for Systematic Reviews and Meta-Analyses (PRISMA) method proposed by ~\citet{page2021prisma}, we conduct the systematic literature review (SLR).

\subsection{Literature Searching Methods}
Before initiating the systematic research collection, we carried out an unstructured exploration of some commonly used terms and ideas concerning the topic. Keywords such as "wastewater", "epidemiology", "WBE", "COVID-19/SARS-CoV-2", "analysis", "modeling", and "surveillance" were commonly used to identify records of peer-reviewed articles in the multidisciplinary literature.

Based on the previous research review, we chose the following databases for our search:
\begin{itemize}
	\item Web of Science: www.webofscience.com
	\item Scopus: www.scopus.com
	\item Engineering Village: www.engineeringvillage.com
	\item PubMed: www.pubmed.ncbi.nlm.nih.gov
\end{itemize}

We divided the wastewater-based COVID-19 surveillance problem into two phases for a full-extent study: data collection and data analysis. The data collection part includes sampling, data acquisition and pre-processing, quantification and normalization, etc. The data analysis part includes the analytical models for the WBE data.
We conducted two separate searches on the database, each using different search phrases as indicated in Table \ref{tab:lss}. Such search settings are designed to return as many topic-related results as possible. After that, we filtered and de-duplicated the results to do further analysis.

\begin{table*}
\caption{Literature searching settings.}
\label{tab:lss}
\begin{tabular}{lll}
\hline
                                 & Boolean operator & Keywords                                                         \\ \hline
\multirow{6}{*}{Data Collection} &                  & COVID OR SARS-CoV-2                                \\
                                 & AND              & wastewater sample          \\
                                 & AND              & virus OR viral           \\
                                 & AND              & sampl* OR  detect*     \\
                                 & AND              & RNA OR ribonucleic acid OR genet*  \\
                                 & AND              & "procedure" OR "protocol" OR method* OR quantif* OR estimat* OR measur*       \\ 
                                 & AND              & *pcr*          \\

\hline
                                 
\multirow{5}{*}{Data Analysis}   &                  & COVID OR SARS-CoV-2                  \\
                                 & AND              & "wastewater-based epidemiology" or WBE \\
                                 & AND              & surveill* OR monitor* OR track*          \\
                                 & AND              & predict* OR forecast* OR foreshadow*  OR trend \\
                                 & AND              & analy* OR statisti* OR model* OR correlate* OR relation*  \\ 
\hline

\end{tabular}
\end{table*}

\subsection{Filtering and Selection}
\begin{figure*}[!htb]
    \centering
    \includegraphics[width=0.9\textwidth]{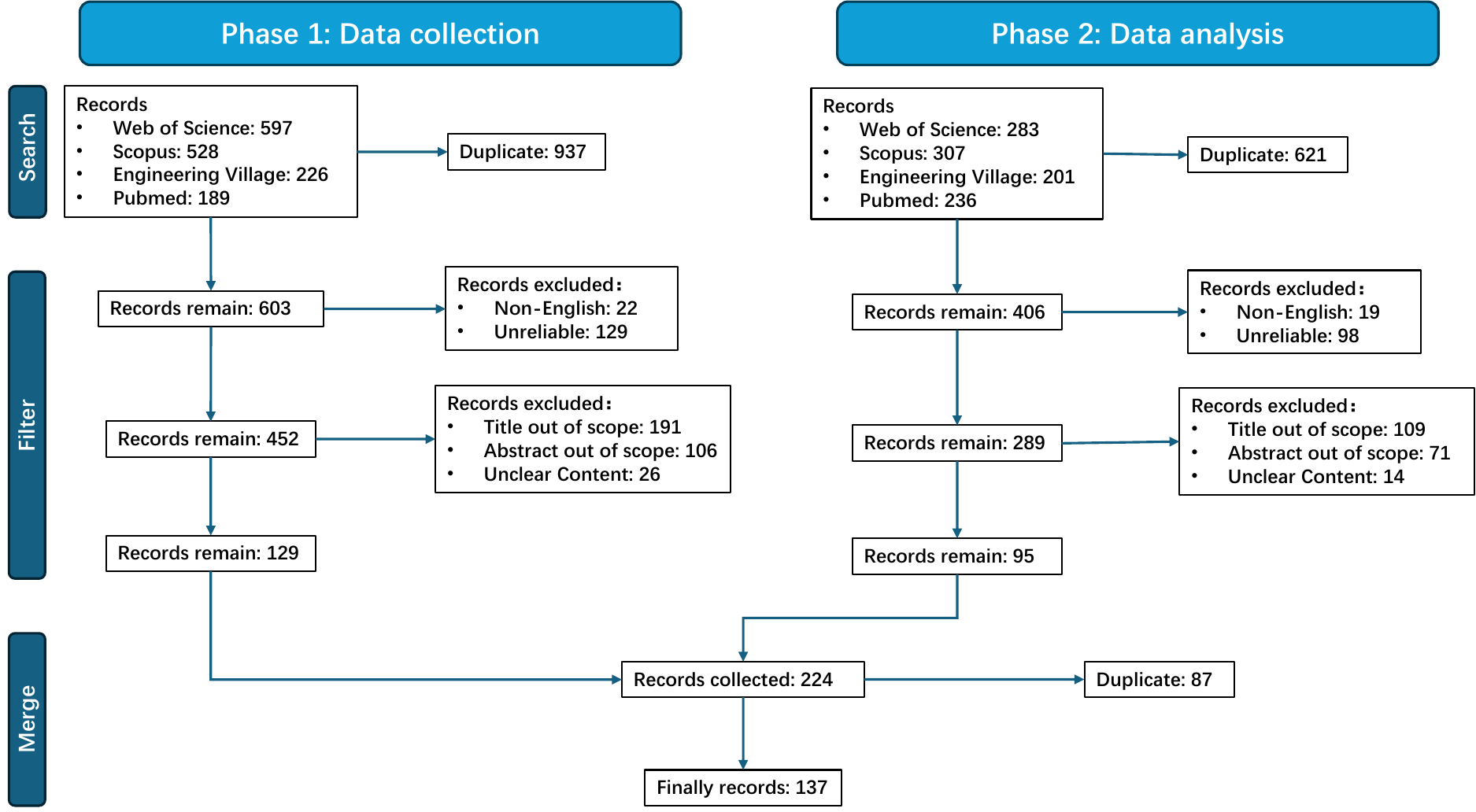}
    \caption{Publication selection process: PRISMA-based flowchart.}
    \label{fig:lss}
\end{figure*}

The criteria provided in Table \ref{tab:lss} were used to search the four literature databases in the time range from January 2020 to October 2023.
From an initial total of 2,567 discovered literature records, we first utilized Zotero to detect duplicate records individually for each search. Next, we applied language filters to choose only English publications.
To ensure the reliability of the selected studies, we only keep original, peer-reviewed articles in our results, leaving out other publishing forms like reviews, short communications, technical reports, letters, notes, abstracts, and surveys. Any work that was accessible yet unpublished was also disqualified. 
After that, we screened the content and selected the literature that is closely related to the perspective of this study based on the article titles and abstracts, and finally, we got 137 related articles. The detailed process and outcome are shown in Figure \ref{fig:lss}.

\hide{
\subsection{Data Summary}

Based on the summary of the literature selected above, we carried out this review work. 

In particular, we focused on the correlation between viral data and clinical data. We selected the relevant literature with good quality and used a MS Excel spreadsheet to include the common information reported. The columns are organized as: Location, Site and population, sample method and frequency, study period, total samples, correlation type, correlation variables, correlation strength, time lag and references.

%
%
%
%
%
%
("COVID-19" OR "SARS-CoV-2" OR "coronavirus") 
AND (wastewater OR "wastewater effluent" OR "wastewater treatment") 
AND (RNA OR "viral RNA" OR "nucleic acid") 
AND (detect* OR quantif* OR "virus detection" OR "viral quantification" OR "viral load")
AND ("quantitative PCR" OR "RT-qPCR")
%
}

\section{Wastewater Sampling}
\noindent Sampling is a critical step for wastewater-based COVID-19 surveillance, which defines the surveillance scope for the disease. In particular, sampling through the sewage can effectively monitor the viral level at a community level or building level; while sampling at the wastewater treatment plant can estimate the infection level at the sewershed level. In addition to the sample location, sample frequency, sample type, and sample method may also affect the effectiveness of disease surveillance and prevalence estimation. This section summarizes the key findings for the above three aspects of wastewater sampling.

\noindent\textbf{Sample Frequency.} WBE is an important tool in monitoring the prevalence of SARS-CoV-2 in the community. Depending on the goal of surveillance, sampling frequency can vary. To screen for the presence of the virus, sampling once per week may be sufficient. To identify infection trends, at least three sampling points within a trend period of interest are needed. The National Wastewater Surveillance System (NWSS) suggests using a 15-day surveillance window for trend reporting~\citep{cdcwastewater}.

\noindent\textbf{Sample Type.} 
Wastewater can be sampled both from sewer systems such as sewage networks and wastewater treatment facilities, and non-sewered systems such as rivers and canals. As the virus concentration in non-sewered systems can be strongly affected by environmental factors, it is hard to make meaningful comparisons between samples collected under different environmental conditions. Therefore, most of the existing wastewater studies tend to focus on samples collected from treatment facilities as shown in Table~\ref{tab:corr_table}.
The wastewater samples collected from sewer systems can be categorized into two different types: (1) untreated wastewater from upstream sewage networks like manholes or treatment plant influent, and (2) treated wastewater from primary sludge in the treatment plant after the first solids removal stage. The advantage of using untreated wastewater from the upstream network or influents is that it can reflect fine-grained viral levels in targeted communities~\citep{layton2022evaluation,cohen2022subsewershed, rondeau2023building}. However, most untreated wastewater samples need to be concentrated prior to viral extraction. For the treated wastewater samples from primary sludge, the concentration step can be eliminated but the viral level in the sample can only be used to evaluate the disease prevalence in the entire sewershed. 

\noindent\textbf{Sample Method.} To collect wastewater samples, there are three commonly used methods: grab, composite and passive sampling. The grab method collects a fixed amount of wastewater at a certain time. The composite method collectively pools multiple grab samples over a certain period of time. While the passive sampling method places devices in wastewater streams to accumulate contaminants without the need for continuous manual intervention.

\begin{table*}[]
\caption{Comparison between sampling methods.}
\label{tab:comparesample}
\begin{tabular}{p{3cm}p{6cm}p{6cm}}
\hline
                                    & Advantages                                                                                                 & Limitations                                                                                                                                            \\ \hline
\multirow{2}{*} {\makecell[l] {Grab Sampling \\ (One-time sample)}}      &  \textbullet~ Fast and simple data gathering satisfies the time-sensitive surveillance needs.                                                                   & \textbullet~ Unable to capture temporal variations and may lead to misrepresentation. 
\\
                                   & \textbullet~ Easy to implement due to its simplicity.                                      &                                                                                                                                                        \\ \hline
\multirow{2}{*} {\makecell[l]{Composite Sampling \\ (Pool of samples)}} & \textbullet~ Collecting and averaging multiple samples over a period provide a more representative picture of viral load. & \textbullet~ Complexity of implementation due to its requirement of sophisticated equipment and more human efforts.                                                \\
                   & \textbullet~ Reducing the risk of missing short-term fluctuations leading to more reliable estimates.                  & \textbullet~ The degradation of the virus RNA over time can affect the accuracy of the results.                                                                     \\ \hline
\multirow{2}{*}{\makecell[l]{Passive Sampling \\ (Accumulation)}}  & \textbullet~Less labor-intensive and more cost-effective. Easy to deploy and retrieve, suitable for remote or difficult-to-access locations.                                                               & \textbullet~ Careful calibration and standardization are needed to eliminate the errors caused by the difference in sampling devices and environmental conditions. \\
                                    & \textbullet~ Time-integrated sampling can give a more comprehensive picture of viral prevalence.                        & \textbullet~ The flow rates and other physical factors can result in potential bias.                                                                               \\ \hline
\end{tabular}
\end{table*}

\begin{figure*}
   \centering
   \includegraphics[width=0.9\textwidth]{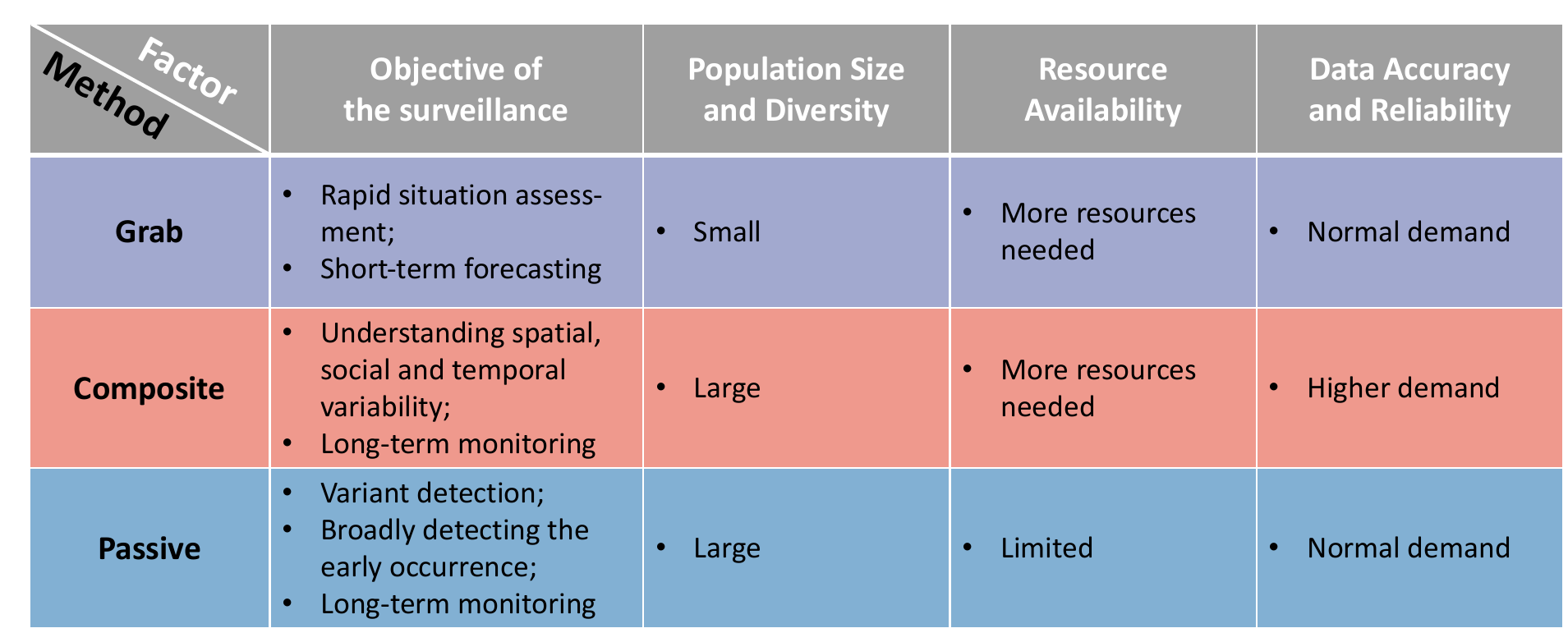}
   \caption{The guidance of sampling method selection. Each column represents a condition to be considered and each box indicates which sampling method we should choose under given circumstances.}
   \label{fig:guidance}
\end{figure*}

In~\citet{gerrity2021early}, a wastewater study in Southern Nevada showed that the SARS-CoV-2 concentration in the composite sample is $10\times$ higher than the early-morning grab samples. 
In~\citet{augusto2022sampling}, a similar study was conducted to evaluate the variability of SARS-CoV-2 RNA concentration in grab and composite samples from both wastewater treatment plants and sewer manholes in Brazil. Their study showed no significant difference between the viral concentrations of the grab and composite samples. 
In particular, the concentrations of composite samples showed greater agreement with concentrations of grab samples collected between 8 a.m. to 10 a.m. The low variability between the two types of samples was also observed in a study at a wastewater treatment plant in Norfolk, Virginia~\citep{curtis2020wastewater}. However, the variability may get amplified when calculating the daily viral load ($\text{viral load}=\text{viral concentration}\times \text{daily influent flow}$) from the viral concentrations. 
Based on previous studies' experience, we summarize the advantages and limitations of different sample methods in Table \ref{tab:comparesample}. Moreover, according to the literature we reviewed, we provide a sampling method selection guideline by considering different factors in real applications as shown in Figure \ref{fig:guidance}.

\section{Sample Testing}
\noindent Sample testing aims to estimate the viral concentration from the wastewater samples, which directly affects the usefulness of downstream data analytic models. Generally, the testing step includes sample pre-processing and virus detection/quantification. To account for the viral loss in the testing step, some lab control methods were introduced to the process. Recent studies suggest that the tested viral concentration should also be normalized with the population served by the sewer system. Correspondingly, different normalization methods were incorporated into the virus quantification model. In this section, we summarize the key advances in sample pre-processing, virus detection and quantification, lab control methods, and normalization methods.

\noindent\textbf{Sample Pre-processing.} The wastewater samples need to be properly processed before being tested. The purpose of sample pre-processing is to remove solids~\citep{jmii2021detection} and inactivate virus/bacteria~\citep{reynolds2022sars}. To remove the solids from the sample, centrifugation, and filtration can be performed. Specifically, the filtration needs to be done with large pore sizes (5$\mu m$ or larger) per CDC's guidance~\citep{cdcwastewater}. In~\citet{yanacc2022detection}, the authors suggested that SARS-CoV-2 RNA might predominate in solids. Therefore, concentration methods focusing on both supernatant and solid fractions may perform better for virus recovery. For the viral inactivation, effective procedures include thermal treatment~\citep{calderon2022monitoring,mcminn2021development}, UV light~\citep{castiglioni2022sars,pellegrinelli2022evaluation} or chemical treatment~\citep{tomasino2021sars}.
Another key step before sample testing is sample concentration, which can help with the detection of SARS-CoV-2 RNA. The concentration step is particularly helpful for untreated wastewater samples as compared to the treated samples as mentioned in the previous section. Effective concentration approaches include ultrafiltration~\citep{dumke2021evaluation, hasing2021comparison}, filtration through electronegative membrane~\citep{barril2021evaluation,jmii2021detection}, centrifugal ultrafiltration~\citep{anderson2021recovery}, ultracentrifugation~\citep{zheng2022comparison}, polyethylene glycol (PEG) precipitation~\citep{alexander2020concentration,farkas2021concentration}, skim milk flocculation~\citep{pino2021detection,philo2021comparison}, and aluminum flocculation~\citep{pino2021detection,salvo2021evaluation}.

\noindent\textbf{Virus Detection and Quantification.} With proper pre-processing and concentration, the wastewater sample is then ready to be tested for SARS-CoV-2 RNA detection and quantification. The key step for the method is to quantify the targeted genetic materials (i.e., SARS-CoV-2 N1, N2 and E genes~\citep{lu2020us, corman2020detection}) with the polymerase chain reaction (PCR). The main step for the PCR test is using special chemicals and enzymes to amplify the targeted genetic materials in cycles. Once the target genes are amplified, they become detectable by lab methods and can be further interpreted to get the viral concentration in the sample. The most common way for RNA detection and quantification is polymerase chain reaction (PCR)-based quantification~\citep{ni2021novel}. In practice, there are different PCR procedures used for SARS-CoV-2 RNA quantification, including RT-LAMP (reverse
transcription loop-mediated isothermal amplification)~\citep{amoah2021rt}, RT-qPCR (reverse transcription-quantitative polymerase chain reaction)~\citep{ahmed2020first}, variations of RT-qPCR~\citep{la2020first,navarro2021sars}, and RT-ddPCR (RT-droplet digital PCR)~\citep{flood2021methods}. In addition to the viral concentration, the number of amplification cycles used to detect the target genes (i.e., the $C_t$ value) can also be used as a criterion to quantify the viral load. Specifically, the lower the $C_t$ value, the greater the amount of viral RNA present in the original sample and vice versa. 

\noindent\textbf{Calibration.} The amount of SARS-CoV-2 virus in the wastewater sample is subject to loss during the sample pre-processing and testing steps. The lost amount may vary by sample quality and testing methods. To assess the lost amount during the process, a frequently used calibration method is matrix recovery control.
A matrix recovery control is a virus that is biologically similar to SARS-CoV-2. Some commonly used control viruses include murine coronavirus (also called murine hepatitis virus), bacteriophage phi6, Pepper Mild Mottle virus (PMMoV), bovine coronavirus, bovine respiratory syncytial virus, and human coronavirus OC43~\citep{ahmed2020comparison, torii2022comparison,hata2020identification,laturner2021evaluating,nagarkar2022sars}. Specifically, the matrix recovery control is spiked into the wastewater sample at a known concentration prior to the pre-processing step. The concentration of the control virus will be tested again after the testing step. The ratio of the virus concentrations before pre-processing and after testing can be used to estimate the recovery rate of the SARS-CoV-2 virus during the entire procedure.
Pa\noindent\textbf{Normalization.} To enable the comparison of viral concentrations across locations and over time, the raw concentrations often need to be normalized by the daily wastewater flow and the population served by the sewer system. As the number of people contributing to the sewershed may vary over time due to factors like tourism and commuting, it is critical to utilize human fecal normalization to account for such changes. 

Human fecal normalization aims to estimate the human fecal content by targeting the organisms that are specific to human feces. Commonly used fecal indicator viral molecular targets include Pepper Mild Mottle virus (PMMoV) and crAssphage~\citep{rosario2009pepper,wilder2021co}. In~\citet{d2021quantitative}, it was shown that PMMoV RNA is relatively stable under different environmental conditions and therefore can boost the correlation between viral signals and COVID-19 cases. The bacterial molecular targets include Bacteroides HF183 and Lachnospiraceae Lachno3~\citep{seurinck2005detection,feng2018human}. 

\noindent\textbf{Quality Control and Quality Assurance}
Quality control (QC) and quality assurance (QA) are critical for ensuring the reliability and accuracy of WBE data for COVID-19 surveillance, which should be employed at every step of the process. Standardized sample collection techniques are the first step in the protocol, which could reduce variability and eliminate contamination. Every sample batch is subjected to stringent quality control measures, such as the use of non-template controls to identify contamination, positive controls to verify assay sensitivity, extraction controls to verify the effectiveness of nucleic acid isolation, and processing blanks to monitor for procedural contamination~\citep{flood2021methods, de2022wastewater, flood2023understanding}. In~\citep{WRFsummit}, the Water Research Foundation (WRF) provided a checklist for QC and QA during the method development process. In particular, the validation of the
assay should include (1) initial precision and recovery controls; (2) matrix spike; (3) estimate of the limit of detection and limit of quantification; and (4) reporting of the equivalent volume of sample analyzed. Once an assay has been developed and validated, the minimally
acceptable QA/QC standards for every assay include (1) detection assay controls; (2) ongoing precision recovery; (3) reporting of the equivalent volume of sample
analyzed; and (4) periodic matrix control spikes.
 On the other hand, to maintain data consistency throughout the analytical phase, routine proficiency testing, calibration, and maintenance should be conducted for all equipments. What is more, comprehensive training programs for personnel, detailed documentation of procedures, and continuous monitoring and evaluation to facilitate ongoing improvements should be included in the QC/QA framework to make sure that the WBE data is reliable and robust.

\section{Data Analytics for Wastewater-based COVID-19 Surveillance }
 
\begin{figure*}[htb!]
    \centering
    \includegraphics[width=0.95\textwidth]{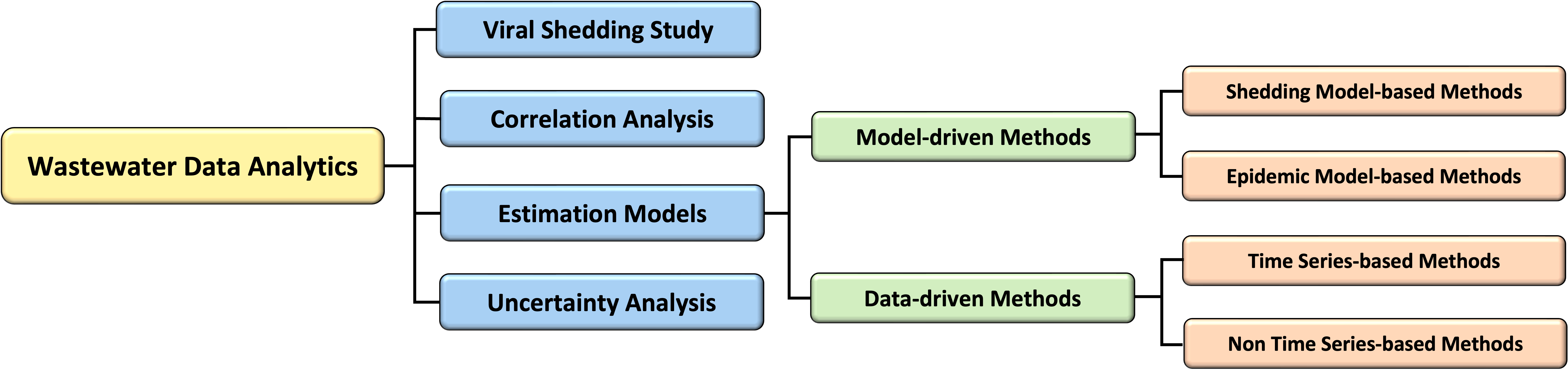}
    \caption{The overview of wastewater data analytics.}
    \label{fig:data_overview}
\end{figure*}

In this section, we review the current literature on wastewater data analytic methods from four perspectives, which include viral shedding studies, correlation analysis, estimation models, and uncertainty analysis. Specifically for the estimation models, we divide the current methods into model-driven methods and data-driven methods. The organization of this section is illustrated in Figure~\ref{fig:data_overview}.

\subsection{Viral Shedding Studies}
The existing viral shedding studies are focused on quantifying the amount of SARS-CoV-2 virus in different types of human waste from infected individuals\footnote{In the context of this paper, "cases" and "infected individuals" have the same meaning and can be used interchangeably.} and the shedding duration of the virus.

\noindent\textbf{Shedding Amount.} \citet{gupta2020persistent} reviewed the literature describing COVID-19 patients tested for fecal virus. The review shows that only 53.9\% of the infected individuals tested for fecal RNA were positive. A more detailed study was conducted in~\citet{jones2020shedding}, which suggests that the SARS-CoV-2 RNA can be detected not only in feces but also occasionally in urine. The likelihood of SARS-CoV-2 being transmitted via feces or urine appears much lower due to the lower relative amounts of virus present in feces/urine. Consequently, the likelihood of infection due to contact with sewage-contaminated water (e.g. swimming, surfing, angling) or food (e.g. salads, shellfish) is extremely low or negligible based on very low abundances and limited environmental survival of SARS-CoV-2. Similar findings were also discovered in~\citet{wolfel2020virological}, where a virological assessment of hospitalized patients with COVID-19 was conducted. Their study indicates that the infectious SARS-CoV-2 virus is exclusively derived from throat or lung samples, but never from blood, urine, or stool samples. 

To calibrate the shedding rate of infected individuals, Schmitz et al. studied the WBE for SARS-CoV-2 by enumerating the asymptomatic COVID-19 cases in a university campus~\citep{schmitz2021enumerating}. The study found that 79.2\% of SARS-CoV-2 infections were asymptomatic and only 20.8\% were symptomatic. To calculate the shedding rate, positive detected cases from the day before, day-of, and four days after sampling were included in the count of infected individuals contributing to viral shedding.  The results showed that the mean fecal shedding rate by the N1 gene was 7.30 ± 0.67 $\log_{10}$ gc/g-feces (log gene copies per gram-feces). 

In addition to the general shedding study on infected individuals, a later study was conducted to explore the association between patient ages and viral shedding amount based on the data from two wastewater sites in Massachusetts~\citep{omori2021age}. Specifically, the viral load in wastewater was modeled as a combination of viruses contributed by different age groups. By incorporating the case count delay, the wastewater viral load was fitted with the daily case count by different age groups. The results indicate that the virus contribution rate of patients from the 80+ yr age group can be ~1.5 times larger than the corresponding rate of patients from the 0–19 yr age group.

\noindent\textbf{Shedding Duration.} A study from ~\citet{gupta2020persistent} suggests that the duration of fecal viral shedding mostly ranges from 1 to 33 days after a negative nasopharyngeal swab. Similar findings were also reported in~\citet{wu2020prolonged}. Moreover,~\citet{wolfel2020virological} reveals that fecal virus shedding peaks in the symptomatic period, and declines in the post-symptomatic phase. ~\citet{miura2021duration} modeled the viral shedding kinetics with the collected data under the Bayesian framework. In particular, the duration of viral shedding and the concentration of virus copies in feces over time are jointly estimated. The results showed that the median concentration of SARS-CoV-2 in feces was 3.4 (95\% CrI\footnote{CrI: Credible Interval.}: 0.24–6.5) log gc/g-feces over the entire shedding period, and the duration of viral shedding is 26.0 days (95\% CrI: 21.7–34.9) from symptom onset date. 

\subsection{Correlation Analysis}
\label{sec:corr-ana}
The correlations between the wastewater viral level and the clinical data (e.g. cases, hospitalization, death) are extensively studied in the current literature. A detailed summarization of correlation studies can be found in Appendix Table~\ref{tab:corr_table}.
To evaluate the correlation between viral data and clinical data, several different correlation metrics are used in the current literature, which include Pearson correlation, $R^2$ for the linear regression model, Spearman's rank correlation, and Kendall's $\tau$ correlation. 
Specifically, the Pearson correlation evaluates the strength and direction of the linear relationship between the clinical data and wastewater level, which can be sensitive to noisy data points or outliers. The $R^2$ is used to illustrate how much of the variance in the dependent variable can be predicted from the independent variables.
Particularly, higher $R^2$ values indicate that the clinical data can be well-fitted by wastewater viral data with a linear relationship.
To relax the correlation from linear constraints, Spearman's rank correlation is used to evaluate the rank consistency between the two data series. 
Similarly, Kendall's $\tau$ correlation is also occasionally used to solve the small sample size problem and the tied values problem, which is defined by the concordance of data pairs. Here, we summarize the key findings from the correlation analysis.

\noindent\textbf{Influential Factors for Correlation.}
The correlation strength between wastewater viral data and clinical data can be affected by many factors. \citet{li2023correlation} conducted a systematic review and meta-analysis on the correlation between SARS-CoV-2 RNA concentration and COVID-19 cases. The review suggested that the correlation coefficients are potentially affected by environmental factors (e.g. temperature, humidity), epidemiological conditions (e.g. vaccination rate, clinical test coverage), WBE sampling design (e.g. sampling method and frequency), and catchment population (e.g. human mobility, demographics of inhabitants)~\citep{li2023correlation, jiang2022artificial,rasero2022associations,kuhn2022predicting, pillay2021monitoring}. In particular, larger variations in air temperature and clinical testing coverage, and the increase of catchment size have strong negative impacts on the correlation between viral concentration and COVID-19 cases. 
The sampling techniques have a negligible impact on the correlation but increasing the sampling frequency can improve the correlation. 
Moreover, extensive correlation studies suggested that the correlation between viral concentration and new cases (either daily new or weekly new cases) is stronger than that of active cases and cumulative cases. 
Also, as the shedding duration of the SARS-CoV-2 virus can be as long as several weeks, the correlations between wastewater viral data and reported cases are often stronger in the pre-peak phase than in the post-peak phase~\citep{roka2021ahead}. 
\noindent Normalizing viral data with fecal indicators can also improve the analysis~\citep{roka2021ahead,scott2021targeted,tandukar2022detection,nagarkar2022sars,d2021catching,d2021quantitative,perez2023long,mohapatra2023wastewater}.
In addition to the aforementioned factors, the availability of home test kits has significantly affected the correlation between wastewater viral data and clinical data. \citet{varkila2023use} analyzed the time series of 268 counties in 22
states from January to September 2022. The study showed that
SARS-CoV-2 wastewater metrics accurately reflected high clinical rates of disease in early 2022, but this association declined over time as home testing increased.

\noindent\textbf{Varying Lag Time.}
Aside from the correlation strength, many existing correlation studies also investigated the lag time between clinical data and wastewater viral load. 
In most cases, the viral load in wastewater is a leading indicator for clinical data, with leading time ranging from 1 day to 2 weeks during peak times considering the time-lag between infection and test confirmation, and asymptomatic infections~\citep{yanacc2022detection,lemaitre2020assessing}. 
However, the viral load may become a lagging indicator during the infection declining phase due to prolonged viral shedding duration~\citep{gerrity2021early}. On the other hand, the lag time for different clinical data types may follow different distributions as well. In general, the lag times of positive tests are shorter than the hospitalization admissions. The lag of hospitalization is further shorter than the death cases~\citep{peccia2020measurement,d2021catching,krivovnakova2021mathematical}.

It is worth mentioning that the lag time may vary significantly by time, location, and catchment population due to the variant accessibility of testing resources and epidemiological conditions of the population~\citep{zhao2022five,kuhn2022predicting,bertels2023time,acosta2022longitudinal,lopez2023predictive,belmonte202320}. 
For example, as the pandemic progressed, many countries have improved their testing infrastructure and reporting systems, leading to more rapid and reliable clinical data. Consequently, the reduction in report delays effectively shortened the lag time observed in WBE studies. However, in the endemic stage, where cases may be underreported due to the availability of home-test kits and reduced report efforts, the correlation between viral loads in wastewater and reported clinical cases has become less robust.

\noindent\textbf{Correlation with Estimated Prevalence.} In some areas, the wastewater viral data was used to estimate the disease prevalence in the sewersheds by leveraging the personal shedding rate and Monte Carlo simulations~\citep{wang2021long,de2022wastewater}. The estimated prevalence was found to be significantly higher than the reported clinical cases in the area due to asymptomatic cases and unreported cases. To thoroughly understand the gap between disease prevalence and reported cases, Layton et al. performed randomized door-to-door nasal swab sampling events in different Oregon communities to infer the community COVID-19 prevalence~\citep{layton2022evaluation}. The estimated prevalence data was then compared with the reported positive cases and the wastewater concentration in the community. Statistical results show that the wastewater viral concentrations were more highly correlated with the estimated community prevalence than with clinically reported cases. Similar results were also observed in~\citet{claro2021long,pillay2021monitoring,gonzalez2021detection,de2022pysewage,saththasivam2021covid}.

\subsection{Estimation Models}
\subsubsection{Model-driven Methods}~\\
\noindent\textbf{Shedding Model-based Methods.}
The key idea for shedding model-based estimation methods is to directly use viral concentration/load and human shedding profiles to estimate the total infected population. The pioneer work was proposed in~\citet{ahmed2020first, ahmed2021sars}, which studied the WBE for SARS-CoV-2 in Australia. In particular, the prevalence of SARS-CoV-2 in the sewershed was estimated using the following formula:
\begin{equation}
    I = \frac{c*f}{p*s}
    \label{eq:model}
\end{equation}
with 
\begin{align}
    &c  = \frac{RNA~copies}{liter~wastewater} &f &=\frac{liter~ wastewater}{day}\\ \nonumber
    &p = \frac{gram~feces}{person~per~day} & s &= \frac{RNA~copies}{gram~feces} \nonumber    
\end{align}
where the infected population $I$ is derived from the viral concentration $c$, wastewater flow rate $f$, fecal production rate $p$, and fecal shedding rate $s$.
The uncertainty and the variability of the independent variables were approximated using a Monte Carlo approach, which yielded a reasonable result that agrees with the clinical observations. Similar analysis was also applied in Brazil~\citep{claro2021long, de2022pysewage}, South Africa~\citep{pillay2021monitoring}, Mexico City~\citep{gonzalez2021detection}, Tehran~\citep{amereh2022association}, Winnipeg~\citep{yanacc2022detection}, Southern Nevada~\citep{gerrity2021early}, Denmark~\citep{nauta2023early}, and Qatar~\citep{saththasivam2021covid}. 
The simple shedding model in Equation~\ref{eq:model} can be easily modified to account for the viral decay~\citep{yanacc2022detection}, the variability on shedding load, the infection-to-confirmation case delay~\citep{fernandez2021wastewater}, and urine viral shedding~\citep{pillay2021monitoring} scenarios.
In~\citet{de2022pysewage}, the model is further developed into a user-friendly web application, pySewage~\citep{pysewage}, to predict the number of infected people based on the detected viral load in wastewater samples, which may be applied to monitor ongoing outbreaks.

\begin{figure*}
    \centering
    \includegraphics[width=0.80\textwidth]{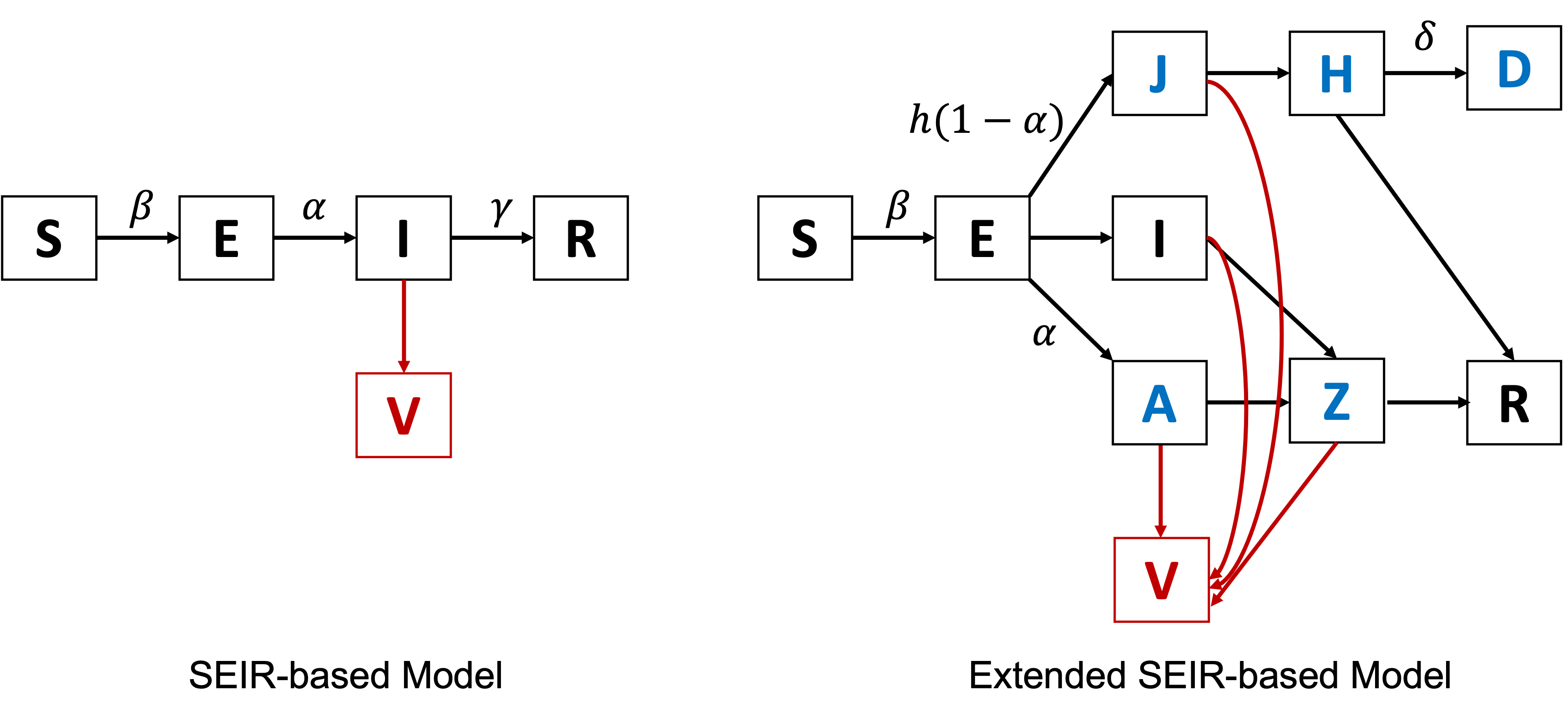}  
    \caption{The SEIR model and extended SEIR model used in the current literature.}
    \label{fig:epimodel}
\end{figure*}

\noindent\textbf{Epidemic Model-based Methods.}
Another line of model-driven methods is to fit the wastewater data into epidemiological models like the susceptible-exposed-infectious-recovered model (SEIR-model~\citep{anderson1979population}) to infer the dynamics of the disease. The framework was first proposed in~\citet{mcmahan2021covid}. Specifically, the framework assumes that the spread of COVID-19 follows the SEIR model and that the viral load in wastewater is solely contributed by the infected population as illustrated in the left panel of Figure~\ref{fig:epimodel}. Let $V_{ij}(t)$ denote the virus shed by individual $i$ on day $t$, who become infected on day $j$, then $V_{ij}(t)$ can modeled by a simple equation below\hide{~\eqref{eq:seir_viral}}
\begin{align}
    V_{ij}(t) = \delta_{ij}&\{10^{\frac{\phi_{ij}(t-j)}{5}}I(j<t\leq 5+j)\\ \nonumber
    ~&+10^{\psi_{ij}^{-\frac{(\phi_{ij}-\psi_{ij})(t-5-j)}{5}}}I(t>5+j)\}  \label{eq:seir_viral}
\end{align}
where $\delta_{ij}$ is the number of grams of feces contributed by the $i$th individual who was infected on the $j$th day, $\phi_{ij}$ is the $\log_{10}$ maximum RNA copies per gram of feces being shed, and $\psi_{ij}$ is the $\log_{10}$ RNA copies per gram of feces being shed 25 days after being infected. To further account for viral decay in the sewage system, a holding time and system temperature-dependent decay model is applied to $V_{ij}$ to approximate the viral loss in the collected samples. The proposed framework was fitted into the wastewater surveillance data in South Carolina from May 2020 to August 2020. The model prediction reveals that the rate of unreported COVID-19 cases was approximately 11 times than that of confirmed cases, which aligns well with the independent estimation of the ascertainment rate in South Carolina.
Following the same framework, McMahan et al. propose to calibrate the SEIR model within a small community for fine-grained analysis~\citep{mcmahan2022predicting}. The study was carried out on a university campus by analyzing the viral RNA copy rates in sewage and the number of SARS-CoV-2 saliva-test-positive individuals among students~\citep{mcmahan2022predicting}.
 A strong correlation was observed between the RNA copy rates and the number of infected individuals. The study also suggested that the most sensitive parameter in calibrating the SEIR model is the maximum shedding rate. Regressing the saliva-test-positive infected individuals on predictions from the SEIR model based on the RNA copy rates yielded a slope of 0.87, which further demonstrated the effectiveness of the proposed framework. In~\citet{phan2023simple}, Phan et al. extended the framework to incorporate the effect of temperature on viral loss into the model. The extended model was tested on the wastewater data in the Greater Boston Area from October 2020 to January 2021. The results showed that the model can successfully recapitulate the temporal dynamics of viral load in wastewater and predicted the true number of cases peaked earlier and higher than the number of reported cases by 6–16 days during the second wave of the pandemic in the area.

Directly inferring the SEIR model from the viral load in wastewater may yield unstable results due to noisy viral fluctuations. To address this issue, some statistical models were explored to reconstruct the epidemic model. \citet{fazli2021wastewater} proposed to utilize the partially observed Markov processes model (POMP~\citep{king2016statistical}) to infer the population in $S, E, I, R$ compartments respectively from the observed viral load and reported cases. Depending on the usage of observed data, three different variants were derived from the framework, which includes "SEIR-VY", "SEIR-V" and "SEIR-Y". Specifically, model "SEIR-VY" uses both viral load and case
counts to fit the parameters, whereas model "SEIR-Y"
and "SEIR-V" utilizes only case counts and viral load,
respectively. The evaluation results demonstrated that a simple SEIR model based on viral load data can reliably predict the number of infections in the near future. Another direction of the study was to use the extended Kalman filter (EKF~\citep{kalman1960new}) to reconstruct the SEIR model~\citep{proverbio2022model}. The proposed framework was used to infer shedding populations, the effective reproduction number, and future epidemic projections. The framework was tested on the wastewater data from different regions. The results showed that the inferred case number is well correlated with the true detected case numbers with correlation coefficients ranging between  0.7 and 0.9. The study also validated that frequent sampling improves the model calibration and the subsequent reconstruction performance. 

 The limitation of the previously mentioned SEIR-based framework is that it assumes all the infected individuals follow the same shedding model. In reality, the shedding models of asymptomatic infections and hospitalized infections may vary significantly from each other. To address this issue, Nourbakhsh el al. presented an extended SEIR model as illustrated in the right panel of Figure~\ref{fig:epimodel}~\citep{nourbakhsh2022wastewater}. Specifically, the infected individuals are further categorized into four subgroups: infection (I), infectious and later admitted to hospital (J), asymptomatic infectious (A), and hospitalized (H). Furthermore, considering some recovered cases may still shedding virus through feces, the recovered group is also divided into two subgroups: non-infectious but still shedding virus (Z) and recovered (R). The model was fitted by the clinical data (both hospitalization and confirmed cases) from three Canadian cities and has provided good estimation on actual prevalence, effective reproduction number, and future incidences. In addition, the model was also used to perform exploratory simulations to quantify the effect of surveillance effectiveness, public health interventions, and vaccination on the discordance between clinical and wastewater data.

 The aforementioned frameworks are predominately based on single-strain epidemic analysis, which cannot effectively deal with the spread dynamics of multiple strains. ~\citet{pell2023emergence} presented a four-dimensional modified SIR model to study disease dynamics when two strains are circulating in the population. The study was applied to understand the emergence of the SARS-CoV-2 Delta variant in the presence of the Alpha variant using the wastewater data from Massachusetts. In the model, a time delay is incorporated to account for temporary cross-immunity induced by the previous infection with the established (or dominant) strain. The study finds that the time delay does not influence the stability of equilibrium and is hence a harmless delay. However, the equilibrium is governed by the basic reproduction numbers of the two strains in nontrivial ways due to the inclusion of cross-immunity.

\subsubsection{Data-driven Methods}~\\
\noindent\textbf{Time Series-based Methods.}
In exploiting the predictive power of the wastewater data from a data-driven perspective, some time series-based methods have demonstrated their effectiveness in short-term forecast tasks. ~\citet{karthikeyan2021high} experimented with the multivariate autoregressive integrated moving average (ARIMA) model to predict the number of new positive cases from the historical case data, wastewater data, and sample collection date in San Diego from July to October 2020. Specifically, the model was used for 1-week to 3-week advance case predictions. To evaluate the model, the Pearson correlation $r$ between the observed cases and predicted cases 
 and the Root Mean Squared Error ($RMSE$) of predicted cases were calculated. For the 1, 2, and 3-week advance forecast tasks, the correlation coefficient and $RMSE$ were $r$ = 0.79, 0.69, and 0.47 and $RMSE$ = 50, 59, and 70, respectively. 

In~\citet{cao2021forecasting}, a vector autoregression (VAR) model was utilized to predict new cases from historical cases and viral concentration in Indiana (PA) from April 2020 to February 2021. The Mean Average Percentage Error (MAPE) for 1-3 week case predictions were 11.85\%, 8.97\% and 21.57\%, respectively. The study suggests that short time series can reliably predict cases 1-week ahead but are not adequate for predicting cases 3 weeks ahead. To improve the robustness of long-term prediction tasks, a longer time series is needed. Moreover, the paper also studied whether different representations of viral data would affect the prediction results. Their study shows that the log-scaled representation of viral concentration has the best interpretation ability of the data, while the original viral concentration has a stronger forecasting ability under the VAR model framework.

The ARIMA model and VAR model were systematically compared in a wastewater surveillance study in Detroit from September 2020 to August 2021~\citep{zhao2022five}. The study showed that the autoregression model with seasonal patterns (SARIMA) and the VAR model are more effective in predicting COVID-19 incidence compared to the ARIMA model. Specifically, the correlation between VAR predicted cases and observed cases is around 0.95 to 0.96 for the 1-week advance forecast task. Similarly, the correlation for the SARIMA-model is around 0.94 to 0.95. While for the ARIMA model, the correlation is only around 0.4 to 0.67.

Another line of time series-based methods is derived from the spatiotemporal methods, which take both spatial information of sewersheds and temporal information of viral load into account in the estimation model. ~\citet{li2023spatio} proposed a spatially continuous statistical model that quantifies the relationship between viral concentration and a collection of covariates including socio-demographics, land cover and virus-associated genomic characteristics at the sewersheds while accounting for spatial and temporal correlation. The model is used to predict the weekly viral concentration at the population-weighted centroid of the 32,844 Lower Super Output Areas (LSOAs) in England, then aggregate these LSOA predictions to the Lower Tier Local Authority level (LTLA). In addition, the model is also used to quantify the probability of change directions (decrease or increase) in viral concentration over short periods.

\begin{figure*}[!htb]
    \centering
    \includegraphics[width=0.9\textwidth]{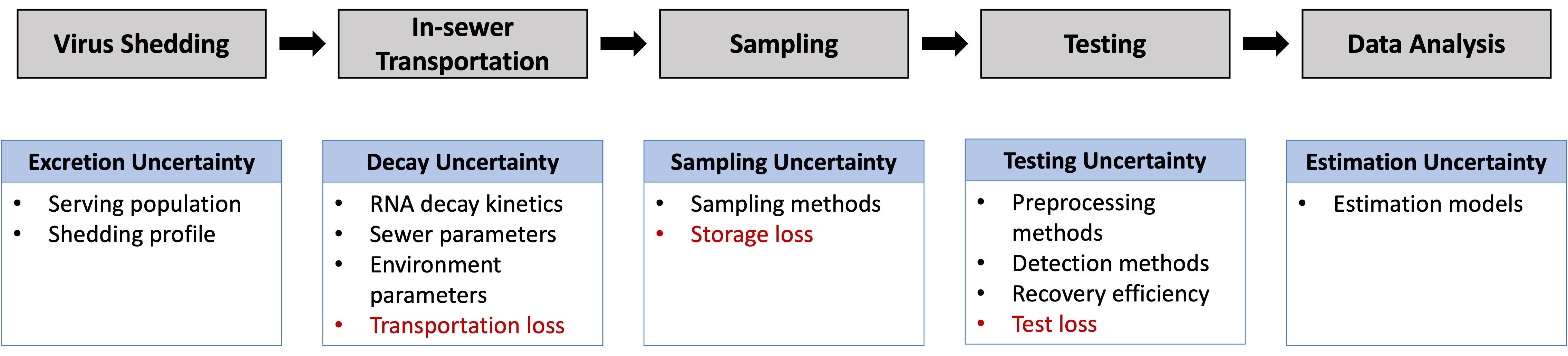}
    \caption{The uncertainty of wastewater-based COVID-19 surveillance.}
    \label{fig:uncertainty}
\end{figure*}

\noindent\textbf{Non-time Series-based Methods.}
A wide range of regression models have been applied to the wastewater data for case prediction due to the ease of implementation and explanability. The simplest regression model assumes that the number of cumulative cases at time $t+\tau$ is linearly related to the viral concentration at time $t$, and has demonstrated its effectiveness for short-term case prediction~\citep{joseph2022assessing}. ~\citet{li2023wastewater} applied the random forest model to predict COVID-19-induced weekly new hospitalizations in 159 counties across 45 states in the United States of America (USA). In particular, different models were established to predict three different hospitalization indicators: weekly new hospitalizations, census inpatient sum, census inpatient average. For each hospitalization indicator, a variety of features, such as Community Vulnerability Index~\citep{smittenaar2021covid}, vaccination coverage, population size, weather, viral concentration, and wastewater temperature, were fed into the model. The study showed that the model can accurately predict the county-level weekly new admissions, allowing a preparation window of 1-4 weeks. In addition, it also suggests updating the training model periodically to ensure accuracy and transferability, with mean absolute error within 4-6 patients/100k population for upcoming weekly new hospitalization numbers.
~\citet{aberi2021quest} compared eight different regression models for COVID-19 surveillance with the wastewater data from four treatment plants in Austria from May to December 2020. The tested models include Linear Regression (LR), Polynomial Regression (PL), $k$ Nearest Neighbor (KNN), Multilayer Perceptron (MLP), Support Vector Regression (SVR), Generalized Additive Models (GAM), Decision Tree (DT) and Random Forest (RF). The study showed that simple models like PL and KNN outperform more complex models such as GAM, SVR, and MLP with slight differences. Similarly, Vallejo et al. applied linear regression, generalized additive model and locally estimated scatterplot smoothing model (LOESS) for COVID case prediction in Northwest Spain~\citep{vallejo2022modeling}. In addition to the wastewater data, some relevant atmospheric variables (e.g. rainfall, humidity, temperature) are also considered in the models. The results showed that the LOESS model yields the least prediction error with $R^2=0.88$. The $R^2$ for the linear and GAM model are $0.85$ and $0.87$, respectively. By changing the prediction period, the study found that the reliability of the model predictions could change by time due to different causes such as the change of SARS–CoV–2 variants. In~\citet{anneser2022modeling}, the linear and the GAM model were compared with Poisson model and Negative Binomial model to predict the cases from the wastewater data in the three New England regions. The models that fit the data best were linear, GAM, and Poisson model with very small differences on $R^2$ and $RMSE$. The same set of models were tested on the wastewater data in Oklahoma city from November 2020 to March 2021, with some sociodemographic factors (e.g. age, race and income) considered in the models~\citep{kuhn2022predicting}. The best results were obtained using a multivariate Poisson model. Consistent with the finding in~\citet{vallejo2022modeling}, the performance of the Poisson model varies by the time of study. Specifically, its accuracy decayed from 92\%, during November 2020 until the end of January 2021, to 59\% during February and March 2021. 
In~\citet{morvan2022analysis}, the shedding model in~\eqref{eq:model} and gradient boosted regression trees (GBRT) were combined to estimate the COVID prevalence in England with the wastewater data from 45 sewage sites. The estimated prevalence was within 1.1\% of the estimates from representative prevalence survey~\citep{morvan2022analysis}. 
In~\citet{xiao2022metrics}, the changing dynamics between the reported cases and wastewater viral load were explicitly studied. Specifically, the clinical reported cases were modeled as the convolution between the scaled wastewater data and an unknown transfer function. It was hypothesized that the transfer function could be fit by a beta distribution. The model was fit into the wastewater surveillance data in the Boston area from March 2020 to May 2021. The results showed that the transfer function has a broad peak and long tail before mid August 2020, indicating that the process of infected individuals getting counted as cases has a broad distribution, with some individuals getting reported very quickly but others taking up to weeks. In this case, wastewater viral load can be used as an early indicator of disease dynamics before clinical test results come back positive. After mid August, the transfer function becomes more sharply peaked, indicating that wastewater and reported cases track each other closely. Consequently, wastewater viral load have less utility as an early warning signal as increased clinical testing capacity effectively captures new infections in a timely manner. 

In addition to the aforementioned simple regression models, some deep learning-based models are also explored for the wastewater-based epidemic surveillance tasks~\citep{zhu2022covid,jiang2022artificial,li2021data,galani2022sars}. Specifically, the artificial neural network model (ANN) and adaptive neuro fuzzy inference system (ANFIS) have proven effective in different studies for case prediction tasks when compared with linear models and random forest~\citep{li2021data}. By incorporating the catchment information, weather, clinical testing coverage, and vaccination rate features into the ANN model, the effective reproduction rate can be estimated as studied in~\citet{jiang2022artificial}.

Aside from the effectiveness of learning models, the features used to feed the learning models may also have an impact on the prediction results. In~\citet{li2021data}, the study indicated that the air and wastewater temperature played a critical role in the prevalence estimation by data-driven models. Also, normalizing and smoothing the wastewater data~\citep{aberi2021quest} or transforming the viral load into log scale~\citep{vallejo2022modeling} can help in fitting the models as well. To better understand the spread of the disease and the effect of public health response, Xiao et al. proposed to monitor the ratio between wastewater viral load and clinical cases (WC-ratio) and the time lag between wastewater and clinical reporting in addition to viral load alone~\citep{xiao2022metrics}. Specifically, when the WC ratio is high, it implies that the existing testing capacity has not kept pace with exponentially rising new cases, which nevertheless are detected in wastewater surveillance. Conversely, a low WC ratio indicates that clinical tests are capturing the majority of infections reflected in wastewater viral load. When this ratio is stable and low, it implies that the existing testing capacity is sufficient to assess the extent of new infections.
The time lag, on the other hand, may reflect the accessibility of test facilities. ~\citet{kuhn2022predicting} showed the lag was significantly lower for areas with a higher household income and a higher proportion of the population aged 65 or older, but higher for areas with a high proportion of Hispanic inhabitants.

\subsection{Uncertainty Analysis}\label{sec:uncertain}
The accuracy of wastewater-based COVID-19 surveillance is limited by the uncertainty and inevitable viral loss introduced in each process step. \citet{wade2022understanding} analyzed different sources of uncertainty, which include (1) serving population, which may change by the population immigration across regions; (2) fecal shedding rate, which varies among individuals and over the infection course; (3) sewage network characteristics, such as the percentage of gravity or pressurized pipes, the size of the network, and retention capacity; (4) sampling strategies; and (5) sample testing methods. In Figure~\ref{fig:uncertainty}, we summarize the possible uncertainties introduced in each key steps of the WBE system.

\citet{li2021uncertainties} systematically studied the uncertainty in estimating SARS-CoV-2 prevalence by WBE. The study suggested that the uncertainty caused by the excretion rate can become limited for the prevalence estimation when the number of infected persons in the catchment area is more than 10. As for the sampling methods, grab sampling contributed the highest uncertainty (around 30\% on average) while a continuous flow-proportional sampling method showed <10\% uncertainty. The uncertainty introduced at the testing stage was the dominant factor. Therefore, it is important to use surrogate viruses as internal or external standards during the virus test process. Overall, WBE can be considered as a reliable complementary surveillance strategy for SARS-CoV-2 with reasonable uncertainty (20–40\%). 

It is worth mentioning that the missing values and outliers in the collected WBE data may also impact the output of estimation models. Unfortunately, the remedy strategies are not frequently discussed in most of the literature. In general, the missing values within a short time interval can be effectively imputed by leveraging the observed nearby data points. Some useful methods include mean imputation, simple interpolation methods (e.g., linear interpolation, spline interpolation), or even some time-series based methods as introduced in~\citet{fang2020time}. While to detect the outliers, we can utilize some well-established time-series abnormally detection methods as introduced in~\citet{shaukat2021review}.

\section{Case Study}
This section presents the application of WBE and its influence on public health initiatives through case studies from different regions. 
Additionally, it discusses the possibility of expanding and maintaining WBE as a regular surveillance technique for COVID-19 and other infectious diseases.

\subsection{NWSS in the U.S.}
In the United States, the CDC led the National Wastewater Surveillance System (NWSS) in September 2020 to respond to the COVID-19 epidemic. CDC developed NWSS to track the presence of SARS-CoV-2 in wastewater samples collected across the country~\citep{cdc2023national}.

The CDC uses the following procedures for surveillance: (1) collect wastewater samples from treatment plants in the sewersheds; (2) send samples to environmental or public health labs for SARS-CoV-2 testing; (3) submit testing data to CDC through the online NWSS Data Collation and Integration for Public Health Event Response (DCIPHER) portal; (4) analyze the reported data with the NWSS DCIPHER system and report the findings to the health department for COVID-19 response; (5) publish the results through the CDC’s COVID Data Tracker.

\begin{figure}
    \centering
    \includegraphics[width=0.45\textwidth]{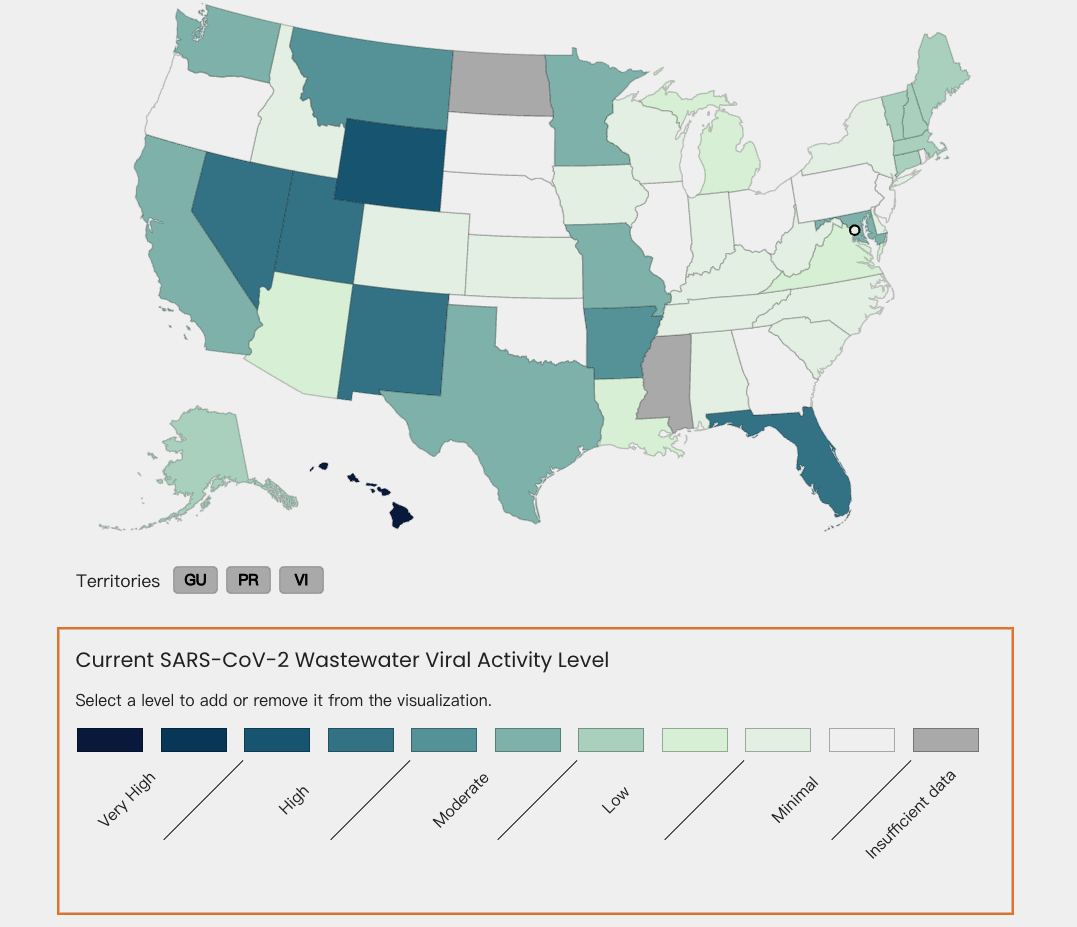}
    \caption{Current Viral Activity Levels Given by ~\citet{cdc2024covid}}
    \label{fig:nwss}
\end{figure}

The NWSS is implemented in all 50 states, 3 territories, and 5 tribal organizations as shown in Figure \ref{fig:nwss}, which provides early detection of changes in disease trends before trends are seen in clinical cases. This information can be used to prepare healthcare providers and hospital systems for upcoming increases in clinical visits, hospitalizations, and demands from other public health prevention efforts. Such wastewater monitoring data is complementary to other public health surveillance data, which enables better variants tracking and outbreak detection. The monitoring system is fast and efficient due to its independence from the medical systems, which makes it able to circumvent the potential delay caused by healthcare accessibility, test availability, and also the incubation period.

\begin{figure}
    \centering
    \includegraphics[width=0.45\textwidth]{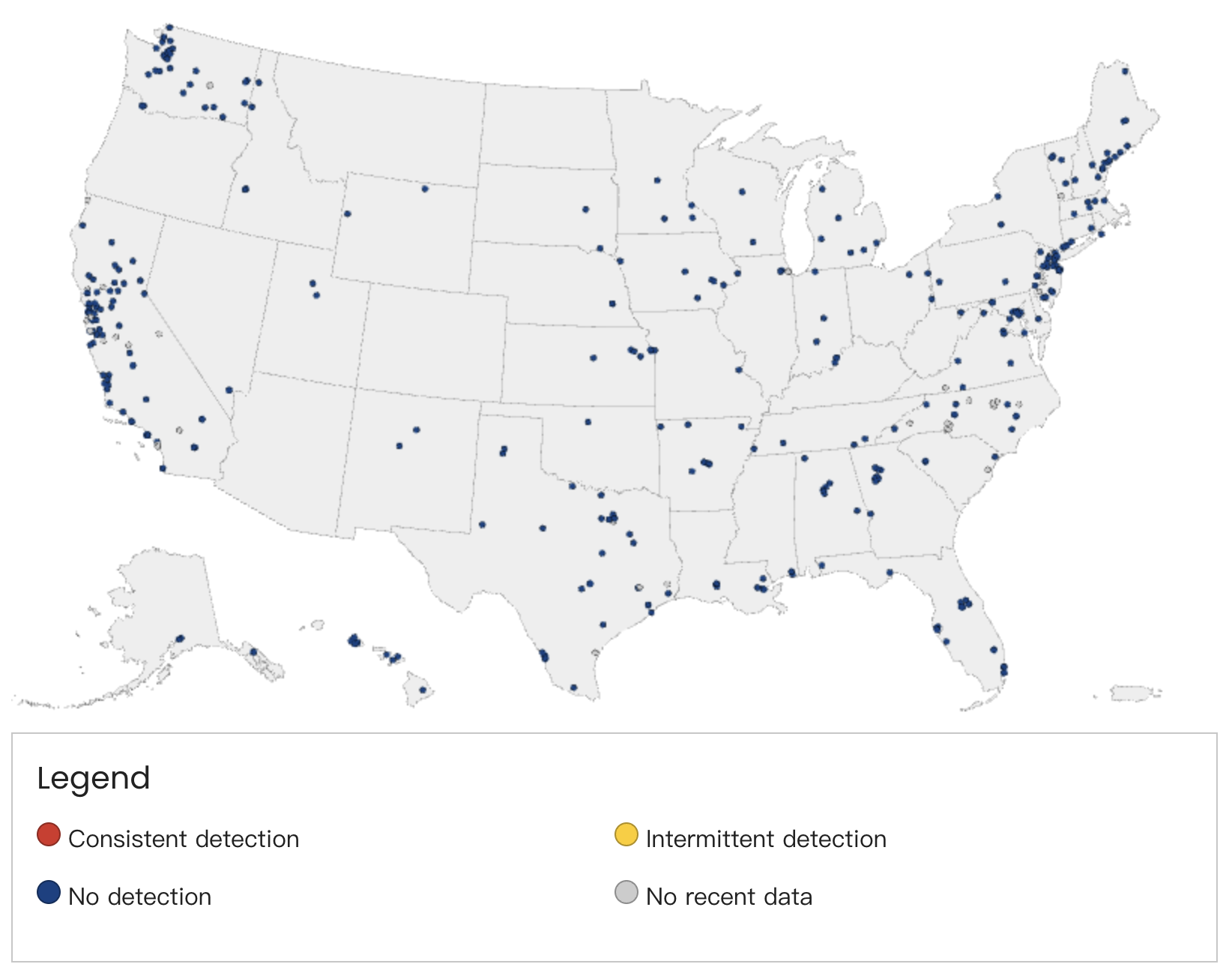}
    \caption{Mpox Virus Detection in Wastewater~\citep{cdc2024mpox}}
    \label{fig:mpox}
\end{figure}

\begin{figure}
    \centering
    \includegraphics[width=0.45\textwidth]{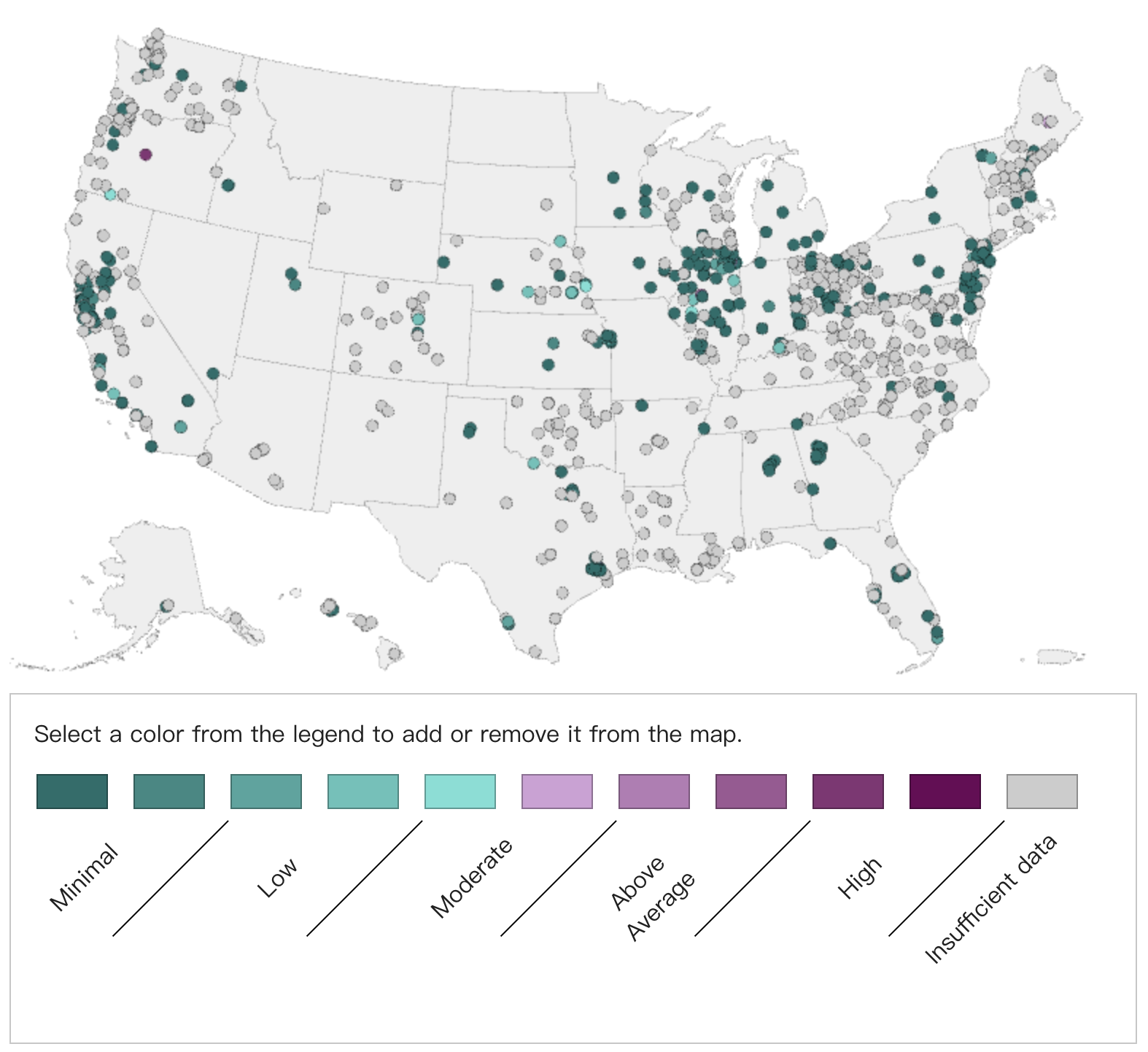}
    \caption{Wastewater Data for Influenza A~\citep{cdc2024influenza}}
    \label{fig:influenzaA}
\end{figure}

In addition to COVID-19, the CDC also uses the NWSS to monitor the Mpox and Influenza A as shown in Figure\ref{fig:mpox} and Figure\ref{fig:influenzaA} ~\citep{cdc2024nwss}. In particular, the Mpox virus detection in wastewater is tracked in a rolling window of 4 weeks. The detection results are classified into 4 categories, which include ``Consistent Detection'' when the Mpox virus was detected in more than 80\% of samples in the past 4 weeks AND the most recent detection was within the past 2 weeks; ``Intermittent Detection'' when the virus was detected in 1\% to 80\% of samples in the past 4 weeks AND the most recent detection was within the past 2 weeks; ``No Detection'' when the virus was not detected in any samples from the site in the past 4 weeks OR the most recent detection was more than 2 weeks ago; and ``No Recent Data'' when too few samples were submitted (fewer than 3) in the past 4 weeks. On the other hand, in the Influenza A monitoring program, sampling sites are categorized based on the current level of influenza A compared to the past levels at the same site during the 2023-2024 influenza season. When influenza A virus levels are at the $80^{th}$ percentile or higher, CDC will start collaborate with relevant partners to understand the factors that could be contributing to these levels.

According to the working experience of the CDC, wastewater surveillance data are most useful when used with other data, such as overall levels of the virus in wastewater, historical wastewater data for that location, geographical context, and clinical cases. These data can be used to refine the wastewater monitoring results and help to gain a more comprehensive analysis. 
The key limitation of the wastewater surveillance system is bounded by the detection limits of the testing methods. In particular, when the viral concentration is lower than the limits of detection, the viral level can not be effectively approximated by the testing methods. Consequently, low levels of infection in a community would not be captured by the wastewater surveillance method. 

\subsection{Surveillance in Other Countries}
Since early 2020, the UK has been analyzing wastewater for SARS-CoV-2 RNA. The Environmental Monitoring for Health Protection (EMHP) SARS-CoV-2 wastewater monitoring program is led by the UK Health Security Agency (UKHSA) and runs in partnership with multiple government agencies, water companies, and universities. 
The program provides coverage for approximately 74\% of the population in England~\citep{english2022}. 
The first detections of the SARS-CoV-2 virus in the UK were made in wastewater samples that were originally taken for polio surveillance. Since then, separate wastewater surveillance initiatives in England, Scotland, and Wales starting in the early summer of 2020 have demonstrated their effectiveness in COVID-19 surveillance
\citep{wade2020wastewater}. As the epidemic began to decline after 2022, the EMHP monitoring program was scaled down and paused at the end of March 2022.

In China, urban sewage surveillance program was added to the Program for Prevention and Control of Novel Coronavirus Infections issued in 2023. Through community sewage monitoring, areas where infected persons may present are identified on time. Possible infected persons can then be traced and found with the help of other surveillance methods. The sewage surveillance at treatment plants 
is therefore used as a key auxiliary method in assessing the epidemiological trends of COVID-19 infections in available areas in China~\citep{china2023}.

In Canada, the government has created a wastewater dashboard that allows people to track and compare COVID-19, Influenza A, Influenza B, and RSV levels over time in some communities in Canada. The data is from 62 sites, representing 49.58\% of the Canadian population~\citep{canada2022covid}. 
In South Africa, the wastewater surveillance is conducted by the National Institute for Communicable Diseases (NICD) to support the government’s response to infectious disease threats, including COVID-19, Meningococcal disease, Typhoid, Shigellosis and Viral Hemorrhagic Fevers (VHF)~\citep{south-africa}. 
In New Zealand, the Institute of Environmental Science and Research (ESR) tested wastewater for the presence of SARS-CoV-2 across the country, which helped the government to identify outbreaks from 2020 to 2022. This surveillance is now used to understand disease trends in communities and to monitor variants. In their surveillance system, most wastewater samples are collected by autosamplers, which greatly reduces human efforts in the process. For places where such passive sampling is not available, grab sampling are used as an alternative~\citep{new-zealand}.

Wastewater-based surveillance of COVID-19 has been successfully used in many countries around the world. Such surveillance methods can be further scaled up to become a routine surveillance tool. To do so, we should make more efforts on (1) developing automated sampling and testing techniques to reduce labor costs; and (2) conducting extensive research on broad-spectrum surveillance methods to monitor multiple virus simultaneously.

\section{Datasets}
This section summarizes the global wastewater datasets that are publicly available in Table~\ref{tab:data}. The data were collected up to September 1, 2024.
For each dataset, the country, data granularity, area covered, time granularity, time span, current status, and corresponding website are listed.  The data granularity represents the aggregation level of wastewater signals, which could range from building-level to country-level. The area covered column shows the monitoring area of the dataset. Time granularity is used to indicate the sampling frequency of the wastewater data. Specifically, for the datasets labeled with `>1/week', more than one data point were observed in one week overall, but the actual weekly samples may vary along the course. The time span specifies the sampling period of the dataset, while the 'live' column indicates whether the data on the website is still getting updated or not. Lastly, the 'website' column gives the link to the dataset.

\section{Future Directions and Challenges}
Wastewater-based epidemiology has been used as an effective tool to complement conventional clinical testing methods for COVID-19 and other infectious disease surveillance. Although substantial efforts have been made in the area, there are still many challenges to be addressed in future research. These important problems that are worth exploring are identified as follows.

\noindent\textbf{Shedding Variability.}
 Current studies predominantly assume that the infected individuals follow a uniform shedding model with only a few works to account for the variability of the shedding profile. In fact, the shedding amount and duration of SARS-CoV-2 in feces can vary widely between individuals and over time. Factors such as the stage of infection, disease severity, vaccination condition, and individual health condition may all affect the shedding profile. As the shedding model is often directly used to estimate the disease prevalence together with the total viral load in the wastewater, it is therefore crucial to construct customized shedding profiles for different infected individuals.

\noindent\textbf{Sample Testing and Virus Quantification.}
Long-term wastewater-based COVID-19 surveillance is an economical way to detect the outbreak of disease and emerging variants~\citep{karthikeyan2022wastewater,lamba2022sars}. One critical problem for the surveillance systems is the allocation of test resources. To be specific, given a limited budget for sample test resources, it is important to choose the sampling locations and frequency by considering the catchment size, and serving population in the area so that potential outbreaks can be detected as early as possible. 
On the other hand, the virus quantified in the wastewater sample may not reflect the actual amount of virus entering the sewage system because of the limited sensitivity of lab methods and viral decay in the sewage system. Therefore, it is crucial to improve the lab testing methods and understand the virus decay model under various environmental parameters (e.g., temperature, wastewater pH, etc.). In addition, it is essential to extend the current COVID-19 surveillance framework to other infectious diseases, so that it can be reused as a general disease monitoring and early warning system.

\noindent\textbf{Data Analytics.}
The majority of the existing literature takes the wastewater data as a standalone signal for epidemic analysis from site to site, while little effort has been made to study the wastewater data from multiple sites collectively for spatial-temporal pattern analysis. Compared to the standalone analysis, the spatial-temporal analysis is more useful for reconstructing the epidemic process at a panoramic scale. In particular, for some large metropolitan areas that can be divided into multiple sewersheds, local residents may contribute to different sites due to the commute from residential areas to commercial areas. In this way, it is hard to recover the disease spread process without tackling the interdependency across sites. The main obstacle to this research direction is the comparability of the data from different sites. Specifically, the sample collection methods, testing methods, and sewage system structure may vary by site. Correspondingly, the same viral load from different sewersheds may represent different epidemic conditions in reality. To this end, how to effectively compile those data into a uniform framework can be a challenging task to address. Moreover, the uncertainties introduced in the wastewater analytic pipeline are not negligible as explained in Section~\ref{sec:uncertain}. Therefore, it is important to quantify the uncertainty together with the prediction results to ensure the reliability of the results. 

\noindent\textbf{Multimodality Data Integration}
The accuracy and timeliness of COVID-19 prevalence estimates can be greatly improved by integrating WBE data with other modality data, such as clinical test data, mobility data, demographic data, and environmental data~\citep{hopkins2023citywide}.
Numerous previous studies have demonstrated a strong correlation between WBE data and clinical data. Based on this correlation and the rapidity of WBE, wastewater monitoring data is often used as an early warning signal~\citep{d2021catching}.
In turn, clinical data and hospitalization data can be used as truth values to correct the analytical model of WBE. As the model continues to get corrected, the predictions would become more accurate.
Another factor affecting the accuracy of surveillance is population distribution, as the serving population size can have a direct impact on the wastewater viral level. 
In addition, static demographic information, including age, sex, and socioeconomic status, can also help contextualize the WBE data.
On the other hand, population movements across sewersheds should also be considered in the analytical model as they could affect regional population size and thus viral loads. In this scenario, mobility data from mobile phones and transportation systems can provide insights into population movement patterns, helping to identify potential hotspots 
and transmission routes. Integrating this data with  WBE can therefore enhance the understanding of viral spread within and between different communities.

{\clearpage
\onecolumn
\pagestyle{empty}
\begin{landscape}
\scriptsize
\begin{longtable}{ccccccc}
\caption{Current Wastewater Publicly Available Datasets for COVID-19 Surveillance.}\label{tab:data}\\
\textbf{Country}                                                                & \textbf{Data Granularity} & \textbf{Area Covered}                                         & \textbf{Time Granularity}     & \textbf{Time Span}           & \textbf{Live} & \textbf{Websites}                                                                                                                                                                                                                                                                        \\ \hline
USA                                                                    & \begin{tabular}[c]{@{}c@{}}Country/Region/\\ County\end{tabular}   & \begin{tabular}[c]{@{}c@{}}Nationwide \\ (240 counties)\end{tabular}            & Weekly              & Jan 2020 - Current   & Yes  & {\color[HTML]{0000EE} {\ul https://biobot.io/data/}}                                                                                                                                                                                                                         \\ \hline
USA                                                                    & \begin{tabular}[c]{@{}c@{}}County/State/\\ WWTP\end{tabular}       & Nationwide                                                                      & Weekly              & Aug 2022 - Current   & Yes  & {\color[HTML]{0000EE} {\ul https://covid.cdc.gov/covid-data-tracker/\#wastewater-surveillance}}                                                                                                                                                                              \\
\hline
\end{longtable}
\end{landscape}
\twocolumn
}
\noindent\textbf{Ethical Concerns. }
WBE data may pose ethical or even legal concerns if not implemented properly. Although WBE data typically cannot identify individuals, when combined with other data sources, there are potential risks for privacy breaches. Therefore, ensuring anonymization and secure handling of data is essential to protect individual privacy. 
On the other hand, it is necessary to communicate with the public about how WBE data is used, its benefits for public health, and measures to protect privacy, so that implementation of WBE is well accepted by the general public. 
Lastly, developing legal frameworks that govern the use of WBE data is important to address potential liabilities and ensure ethical use. Effective measures along this direction include regulations on data collection, sharing, and usage rights.



\section{Conclusion}
Wastewater-based epidemiology has been demonstrated as a powerful tool for COVID-19 surveillance and trend projection within communities. \hide{During the pandemic of COVID-19, extensive studies have been conducted on different aspects of WBE techniques. Moreover, a myriad of analytical models were applied to various communities of different sizes.}This survey summarizes the wastewater sampling techniques, sample testing methods, data analytical models, and the existing wastewater datasets at a global level. In particular, this survey provides a new taxonomy of data analytical models to help the researcher and practitioner form a systematic view of the area. Most importantly, the reviewed data analytical models can be easily generalized to many other infectious diseases, which can be referred to as guidance to build general disease surveillance systems. Moreover, the comprehensive wastewater datasets at different granularity can serve as a benchmark for validating new surveillance models at various scales. Last but not least, the challenges in the area are discussed, which may help inspire researchers in their future research directions.

\noindent
\textbf{Acknowledgements.} \
We thank members of the Biocomplexity COVID-19 Response Team and the Network Systems
Science and Advanced Computing (NSSAC) Division of the University of Virginia for their thoughtful
comments and suggestions related to epidemic modeling and response support. We also thank scientists at the Virginia Department of Health (VDH)
and the Division of Consolidated Laboratory Services (DCLS) for their collaboration.
This work was partially supported by University of Virginia Strategic Investment Fund award number SIF160, National Institutes of Health (NIH) Grant 1R01GM109718,  NSF Expeditions in Computing Grant CCF-1918656, VDH Contract UVABIO610-GY23, NSF Grant CCF-1908308, PGCoE CDC-RFA-CK22-2204, VDH Contract UVABIO610-GY23.
Any opinions, findings, conclusions, or recommendations expressed in this publication are those of the authors and do not necessarily reflect the views of the funding agencies.
This journal article was supported by the Office of Advanced Molecular Detection, Centers for Disease Control and Prevention through Cooperative Agreement Number CK22-2204. Its contents are solely the responsibility of the authors and do not necessarily represent the official views of the Centers for Disease Control and Prevention.


\bibliographystyle{cas-model2-names}
\bibliography{sample-base}

{\clearpage
\onecolumn
\pagestyle{empty}
\begin{landscape}
\section*{Appendix A}
    \scriptsize
\begin{longtable}{ccccccccccc}
        \caption{Summarization of Correlation Studies between Viral Data and Clinical Data.}\label{tab:corr_table} \\
        \multirow{2}{*}{\textbf{Location}}                                                                      & \multicolumn{4}{c}{\textbf{Sampling}}                                                                                                                                                                                                                                                                                                                                                                                                                                                                                    & \multicolumn{5}{c}{\textbf{Correlation}}                                                                                                                                                                                                                                                                                                                                                                                                                                                                                                                                    & \multirow{2}{*}{\textbf{Ref.}}                      \\
\hline
        
        \end{longtable} 
\end{landscape}
Table~\ref{tab:corr_table} summarizes the correlation studies by their study location, sampling information (i.e., sampling site, method, frequency, and sampling period), and correlation details (i.e., correlation types, correlation variables, correlation strength, and time lag between the two variables). Specifically, the `Var. 2 lag' column represents the lag of clinical data (i.e., $variable$ 2) to viral levels (i.e., $variable$ 1). Therefore, a negative lag time means the corresponding clinical data is leading the wastewater viral data. A positive lag time means the clinical data is lagging the viral data.

\twocolumn

}



\begin{table}[!htb]
    \centering
    \section*{Appendix B}
    \caption{Glossary of Terms}
    \begin{tabular}{lp{5.5cm}}
    \toprule
    \textbf{Name}       & \textbf{Description}                                                                                   \\ \midrule
    ANN                 & Artificial Neural Network model                                                                        \\ 
    ARIMA               & AutoRegressive Integrated Moving Average                                                                \\ 
    LOESS               & Locally Estimated Scatterplot Smoothing model                                                           \\ 
    LSOAs               & Lower Super Output Areas                                                                                \\ 
    NWSS                & National Wastewater Surveillance System                                                                 \\ 
    PCR                 & Polymerase Chain Reaction                                                                               \\ 
    PMMoV               & Pepper Mild Mottle virus                                                                                \\ 
    QA                  & Quality Assurance                                                                                       \\ 
    QC                  & Quality Control                                                                                         \\ 
    RMSE               & Root Mean Squared Error                                                                                 \\ 
    SARIMA              & AutoRegression Model with Seasonal Patterns                                                             \\ 
    SEIR model          & Susceptible Exposed Infectious Recovered model                                                          \\ 
    VAR                 & Vector Autoregression                                                                                   \\ 
    WBE                 & Wastewater-Based Epidemiology                                                                           \\ 
    WC ratio            & the Ratio between Wastewater Viral Load and Clinical Cases                                              \\ 
    \bottomrule
    \end{tabular}\label{tab:acronyms}

    \end{table}

\section*{Appendix C}
\textbf{Correlation Metrics Details}

This section will introduce the details of correlation metrics mentioned in section \ref{sec:corr-ana}. Assume that the time series for wastewater viral data and clinical data are $X=\{x_1,x_2,\ldots,x_n\}$ and $Y=\{y_1,y_2,\ldots,y_n\}$ where the data pairs $(x_t,y_t)$ are aligned at timestamp $t$. The correlations between the two time series under different metrics are defined as follows:

\noindent \textit{Pearson correlation}: the Pearson correlation $r_{XY}$ between time series $X$ and $Y$ is defined as 
\begin{equation}
    r_{XY} =\frac{\sum_{i=1}^n(x_i-\bar{x})(y_i-\bar{y})}{\sqrt{\sum_{i=1}^n(x_i-\bar{x})^2}\sqrt{\sum_{i=1}^n(y_i-\bar{y})^2}} 
\end{equation}
where $\bar{x}=\frac{1}{n}\sum_{i=1}^n x_i$ and $\bar{y}=\frac{1}{n}\sum_{i=1}^n y_i$ are the mean of the two series. The correlation score has a value between -1 and 1, which reflects the linear correlation of variables. One practical problem of Pearson correlation is its sensitivity to noise and outliers.

\noindent\textit{$R^2$ for linear regression model}: assume that the clinical data $Y$ can be fitted by the wastewater viral data $X$ with linear regression model (e.g., $\hat{y_i}=a+bx_i$), then the coefficient of determination $R^2$ can be calculated as
\begin{equation}
    R^2 = 1-\frac{\sum_{i=1}^n(y_i-\hat{y_i})^2}{\sum_{i=1}^n(y_i-\bar{y})^2}
\end{equation} 
The $R^2$ for the linear regression model can be used as a complementary metric for Pearson correlation as it provides a clear interpretation in terms of variance explained by the model. Moreover, it can be extended to multiple regression scenarios where multiple sources of clinical data are integrated into the regression model, which can improve the robustness of the model under noisy settings.

\noindent\textit{Spearman's rank correlation}: The Spearman's rank correlation is used to evaluate the rank consistency between two data series. To calculate the correlation between $X$ and $Y$, the two time series need to be converted into series of ranks $R_X$ and $R_Y$. The correlation coefficient would then be calculated as the Pearson correlation between $R_X$ and $R_Y$. The advantage of Spearman's correlation is that $X$ and $Y$ can be related by any monotonic function rather than the linear correlation as in Pearson correlation.

\noindent\textit{Kendall's $\tau$ correlation}: The Kendall's $\tau$ correlation is defined by the concordance of data pairs. Specifically, for any pairs of data $(x_i,y_i)$ and $(x_j,y_j)$, the two pairs are considered concordant if the sort order of $(x_i,x_j)$ and $(y_i,y_j)$ agrees. Based on that, the correlation can be calculated as
\vspace{-1mm}
\begin{equation}
    \tau =  \frac{\#\text{concordant pairs}-\#\text{disconcordant pairs}}{\text{total pairs}}
\end{equation}\vspace{-1mm}
The Kendall's $\tau$ correlation is similar to Spearman's rank correlation but is generally preferred when the sample size is small and when there are many tied values in the time series.


\end{document}